\documentclass[manuscript]{aastex}
\usepackage{graphicx}
\usepackage{txfonts}
\usepackage{natbib}
\bibpunct{(}{)}{;}{a}{}{,}
\usepackage{epstopdf}

\begin{document}

\title{Hydroxyl as a Tracer of $\mathrm{H_2}$ in the Envelope of MBM40}

\author{David L. Cotten and Loris Magnani}
\affil{Department of Physics and Astronomy, University of Georgia, Athens, GA, 30602}
\email{dcotte1@physast.uga.edu, loris@physast.uga.edu}

\author {Elizabeth A. Wennerstrom}
\affil{Wayne State University School of Medicine, Detroit, MI, 48201}

\author{Kevin A. Douglas}
\affil{Department of Physics and Astronomy, University of Calgary, Calgary, Alberta}

\author{Joseph S. Onello}
\affil{Department of Mathematical Sciences, The University of Montana, Missoula, MT, 59812-0864}

\begin{abstract} 
We observed 51 positions in the OH 1667 MHz main line transitions in the translucent high latitude cloud MBM40. 
We detected OH emission in 8 out of 8 positions in the molecular core of the cloud and 24 out of 43 in the surrounding, lower extinction envelope and periphery of the cloud. Using a linear relationship between the integrated OH line intensity and E(B-V), we estimate the mass in the core, the envelope, and the periphery of the cloud to be 4, 8, and 5 M$_{\sun}$. As much as a third of the total cloud mass may be found in the periphery (E(B-V) $<$ 0.12 mag) and about a half in the envelope  (0.12 $\le$ E(B-V) $\le$ 0.17 mag). If these results are applicable to other translucent clouds, then the OH 1667 MHz line is an excellent tracer of gas in very low extinction regions and high-sensitivity mapping of the envelopes of translucent molecular clouds may reveal the presence of significant quantities of molecular mass. 
\end{abstract}


\section{Introduction}\label{sec:intro}
Most studies of molecular clouds have focused on the dense core regions where star formation is likely to occur \citep{blitz99,bergin07}. The surrounding, lower density, outer regions are less often studied because of the significantly weaker molecular lines and lack of star forming potential. However, this outer region is key for understanding the atomic-molecular interface in interstellar clouds and, possibly, their origin from the surrounding atomic interstellar medium (ISM). A standard way to categorize the smaller molecular clouds (in contrast to Giant Molecular Clouds which have masses $\ge$ 10$^4$ M$_{\sun}$) was proposed by \citet{vDB88} using visual extinction, $\mathrm{A_V}$, to break up clouds into diffuse ($\mathrm{A_V}$ $<$ 1 mag), translucent (1 $\le$ $\mathrm{A_V}$ $\le$ 5 mag), and dark ($\mathrm{A_V}$ $>$ 5 mag) categories. Translucent molecular clouds are relatively easy to detect and map in the CO(J=1-0) transition and represent the vast majority of the high-latitude molecular clouds~\citep[][hereafter MBM]{mbm85}. In contrast to dark molecular clouds which have dense, opaque cores surrounded by lower density, lower extinction ``envelope'' regions, translucent clouds are mostly ``envelope'', at least in terms of spatial extent. 

Translucent molecular clouds also differ significantly from dense, dark clouds in that translucent clouds are often not gravitationally bound. Consequently, their star-forming capability is either significantly less than that of dark clouds or, perhaps, non-existent~\citep{hearty99,mcgehee}. Moreover, the way they form from the atomic ISM is also likely to be different from dark clouds and Bok globules. Some recent ideas invoke formation mechanisms for translucent clouds via shear flows or in regions where two HI velocity components meet~\citep{shore03,barr10a}. Thus, probing the atomic/molecular transition region is especially important for understanding the origin of translucent clouds. 

Traditionally, the physical properties of molecular clouds are determined mainly by using the CO(1-0) line. Because of the mechanisms that destroy CO, it is possible that a regime of molecular gas exists where CO does not effectively trace the molecular content of the clouds \citep{vDB88,mo93,wan93,grenier05,dt07, barr10b}.  This regime is usually a region of very low dust column density surrounding the denser, more opaque cores, with the traditional threshold for sufficient obscuration to produce readily observable molecular lines in emission set at about $\mathrm{A_V}$\ $\sim$ 1 mag~\citep{vDB88}. However, recent work~\citep{chastain06,barr10a} indicates that the CO(1-0) line can be readily detected down to $\mathrm{A_V}$ $\sim$ 0.3 mag. This is significant because it forces a revision in both models of photodissociation regions (PDRs) and in molecular mass estimates for molecular clouds, and it provides the capability to study spectroscopically the transition region between the molecular cloud and the surrounding HI flows~\citep[see, e.g.][]{wan93,barr10a}.

Although the CO(1-0) transition is, historically, the one most commonly used to trace low-density, low-extinction gas, there are at least four other molecular and atomic species with transitions that have been used to trace gas at the edges of cloud cores in the so-called ``envelope'' region of a molecular cloud:\footnote{In terms of the PDR models of atomic-molecular cloud interfaces, we are talking about the region where the carbon transitions from being primarily in the form of C$\rm{I}$ to that of CO~\citep[see, e.g.,][]{ht97}.} (1) The $\mathrm{^{2}\Pi_{3/2}}$ OH ground state main lines at 1665 and 1667 MHz~\citep{wan93, W81, ll96}; (2) the $\mathrm{^{2}\Pi_{1/2}}$ CH ground state main line at 3335 MHz~\citep{mo93}; (3) the ortho 1$_{10}$ - 1$_{01}$ and para 2$_{20}$ - 2$_{11}$ transitions of C$_{3}$H$_{2}$ at 18.3 and 21.6 GHz \citep{cox89}; (4) the $^{3}$P$_1$ $\rightarrow$ $^{3}$P$_0$ fine structure line of atomic carbon at 492 GHz~\citep{ing97}. All of these transitions have been proposed as potential tracers of H$_2$ in regions where the CO(1-0) line is not detectable~\citep[however, see][who claim that CO(1-0) observations at higher sensitivity than the usual 0.1 - 0.2 K rms can trace molecular gas in regions with $\mathrm{A_V}$ as low as 0.3 mag]{cotten12}. Of the low-density tracers apart from CO, the OH main lines are the easiest to observe; they are readily detectable in translucent molecular clouds \citep{mag88,magsis90,barr10b}, OH is considered a precursor molecule to the formation of CO~\citep[e.g.,][]{bd77}, and thus from chemical considerations may be abundant in regions where the CO/H$_2$ abundance is low. 

Previous CO(1-0) maps of MBM40 have focused primarily on the relatively dense, wishbone-shaped core region~\citep{shore03}. In contrast, we studied the molecular envelope of this translucent cloud where the visual extinction is likely to be 0.3-0.5 mag (see below). CO(1-0) observations of this region require integration times of at least 15-20 minutes per position to reach rms values sufficiently low to detect lines in the envelope of this cloud~\citep[][in preparation]{cotten12}. 
Moreover, the resolution of most millimeter wave, single-dish telescopes in North America is too high to allow systematic mapping of the extended envelope regions at sufficient sensitivity during typical observing runs. Observations of the CH and C$_3$H$_2$ lines described above require very long integrations, and the C$\rm{I}$ 492 GHz transition is difficult to observe from the ground. 
Thus, for these reasons, we surveyed the two OH main-line transitions in the envelope region of the translucent cloud MBM40. 
By using these observations to estimate the mass in the envelope and periphery regions of MBM40 we can compare how much molecular mass is tied up there that is not well-traced by traditional CO observational techniques. We discuss the translucent cloud selected for this study, MBM40, in Section 2 and describe our OH observations in Section 3. In Section 4 we discuss the W(OH) - E(B-V) relationship, the mass of OH in the cloud, and the molecular mass of the core and outer regions. The conclusions are summarized in section 5.

\section{The Translucent Cloud MBM40}\label{sec:cloud}

\begin{figure}
\includegraphics[width=160mm]{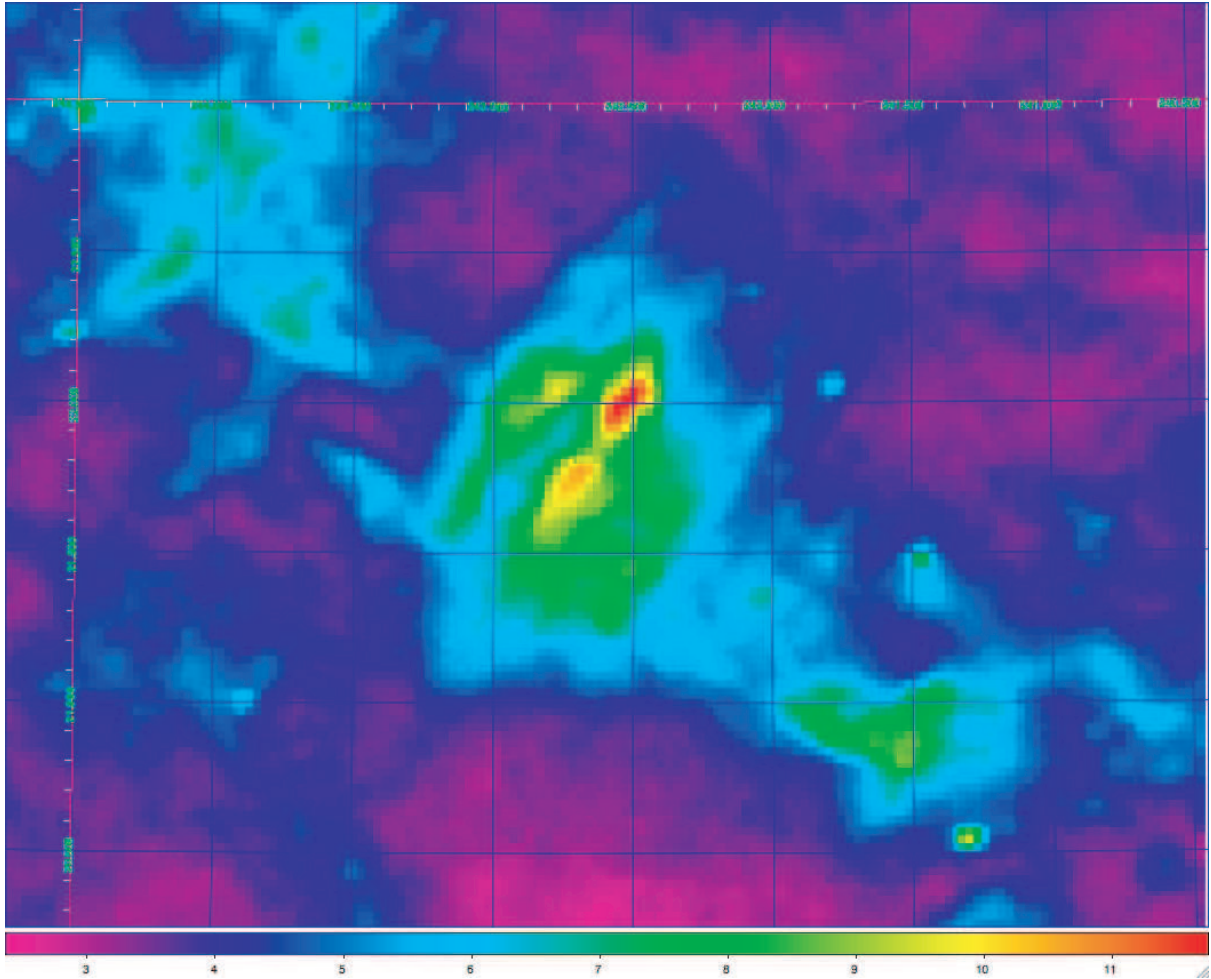}
\caption{IRIS 100 micron emission map of MBM40 and its environs. The central brightest
region shows the horse-shoe or hairpin structure mapped in CO(1-0) by \citet{shore03}. 
The surrounding region is the ``envelope'' of the cloud. The dust structures to the northeast 
and southwest are probably related to MBM40 but have never been mapped in CO. The intensity 
scale is in MJy/ster.}
\label{fig:mbm40-ir}
\end{figure}

MBM40 is a small ($\approx$ 2 square degrees), low extinction ($\mathrm{A_{V}\ \le\ 3\ mag}$)  cloud located at $\ell$ = 37.6\degr\  and $\textit{b}$ = 44.7\degr\  that has been partially mapped in CO,$\mathrm{^{13}}$CO, CS, H$_{2}$CO, CH and H$\rm{I}$ \citep[see][for references]{shore03}. Figure~\ref{fig:mbm40-ir} shows the IRIS$\footnote{IRIS are the new generation of IRAS images which benefit from improved zodiacal light subtraction, calibration, and destriping~\citep{miv05}.}$ 100 micron emission  for a 4\degr \ x\ 4\degr \  region centered on the cloud. 
Clearly, there is significant dust emission surrounding the intense central core region (and associated smaller clouds to the east and west). A detailed CO(1-0) map of the horseshoe-shaped central region of the cloud with an rms sensitivity of 0.7 K  was made by \citet{shore03}. The horseshoe-like shape of this region will be referred to as the ``core'' of the cloud and CO(1-0) contours superposed on an E(B-V) map are shown in Figure 10 of \citet{ccm10}.  Although the \citet{ccm10} CO map effectively traces only the horseshoe or wishbone shaped core region, more sensitive CO integrations reveal CO(1-0) emission in the envelope region surrounding the core of the cloud~\citep[][in preparation]{dev88, cotten12}. We note that \citet{wen07} also demonstrated that the 18 cm OH main lines could be detected from the envelope region of this cloud. 

\citet{chastain06} showed that there is a marked increase in CO(1-0) detections for high-latitude regions with E(B-V) $\ge$ 0.12 mag, so we divided MBM40 into three regions based on the E(B-V) maps by \citet[][hereafter SFD]{SFD98}. 
The three regions: E(B-V) $<$ 0.12 mag, 0.12 $\le$ E(B-V) $\le$ 0.17, and E(B-V) $>$ 0.17, will be referred to as the ``periphery", the ``envelope", and the ``core" respectively, throughout the rest of this paper. 
Of the roughly 2 square degrees that encompass the dust structure associated with MBM40, the periphery, envelope, and core regions cover 1.2, 0.61, and 0.20 square degrees, respectively.

\section{Observations}

Observations of the 18 cm ground state transitions of OH in MBM40 were made using the Robert C. Byrd 100 m radio telescope (hereafter GBT) in Green Bank, WV, during December 2009 and January  2010.\footnote{The GBT is part of The National Radio Astronomy Observatory (NRAO) and is a facility of the National Science Foundation operated under cooperative agreement by Associated Universities, Inc. (www.gb.nrao.edu).} The data were collected with the GBT using position switching with the ``off'' source position chosen to be between 1 and 2 degrees away from the ``on'' position, with exact locations based on the lowest emission region in the SFD-dust maps that allowed for  slewing the telescope in azimuth only.  All scans for a given position were averaged and a second order baseline was subtracted. A Gaussian curve was fit to each line, establishing values for the full width half maximum (FWHM), local standard of rest (LSR) centroid velocity, and peak temperature.

The spectra obtained from the GBT were produced using an autocorrelator spectrometer subdivided into four sections centered on the four 18 cm transitions at 1720.5300, 1667.3590, 1665.4018, and 1612.2310 MHz.  However, given our integration times (typically half an hour per point) the satellite lines were so weak that  they were not detected at all and even the 1665 MHz line had poor signal to noise for most lines of sight. Consequently, we focused our analysis only on the strongest line at 1667 MHz.$\footnote{In thermal equilibrium, the ratio of the 1667:1665:1720:1612 line intensity is 9:5:1:1.}$ For each scan two circular polarizations were measured but were averaged to produce a single spectrum. The bandwidth of each spectrum was 12.5 MHz, which corresponds to a velocity range of 2249 km s$^{-1}$ for the 1667 MHz line, and 2252 km s$^{-1}$ for the 1665 MHz line. The velocity resolution of each channel was 0.059 km s$^{-1}$. At 1667 MHz, the GBT has an angular resolution of 6.2 arc minutes and a beam efficiency of 1.32 times the aperture efficiency which  is defined as the ratio of effective collecting area to physical collecting area, depending on the telescope geometry and on the frequency being observed.  For observations at 18 cm the aperture efficiency is approximately 0.71, which corresponds to a beam efficiency of 0.94 at 18 cm.  The radiation temperature, T$\mathrm{_R}$, is thus T$\mathrm{_A}$/0.94~\citep{Madd09} where T$\mathrm{_A}$ is the antenna temperature.

\begin{figure}
\begin{center}
\includegraphics[width=160mm]{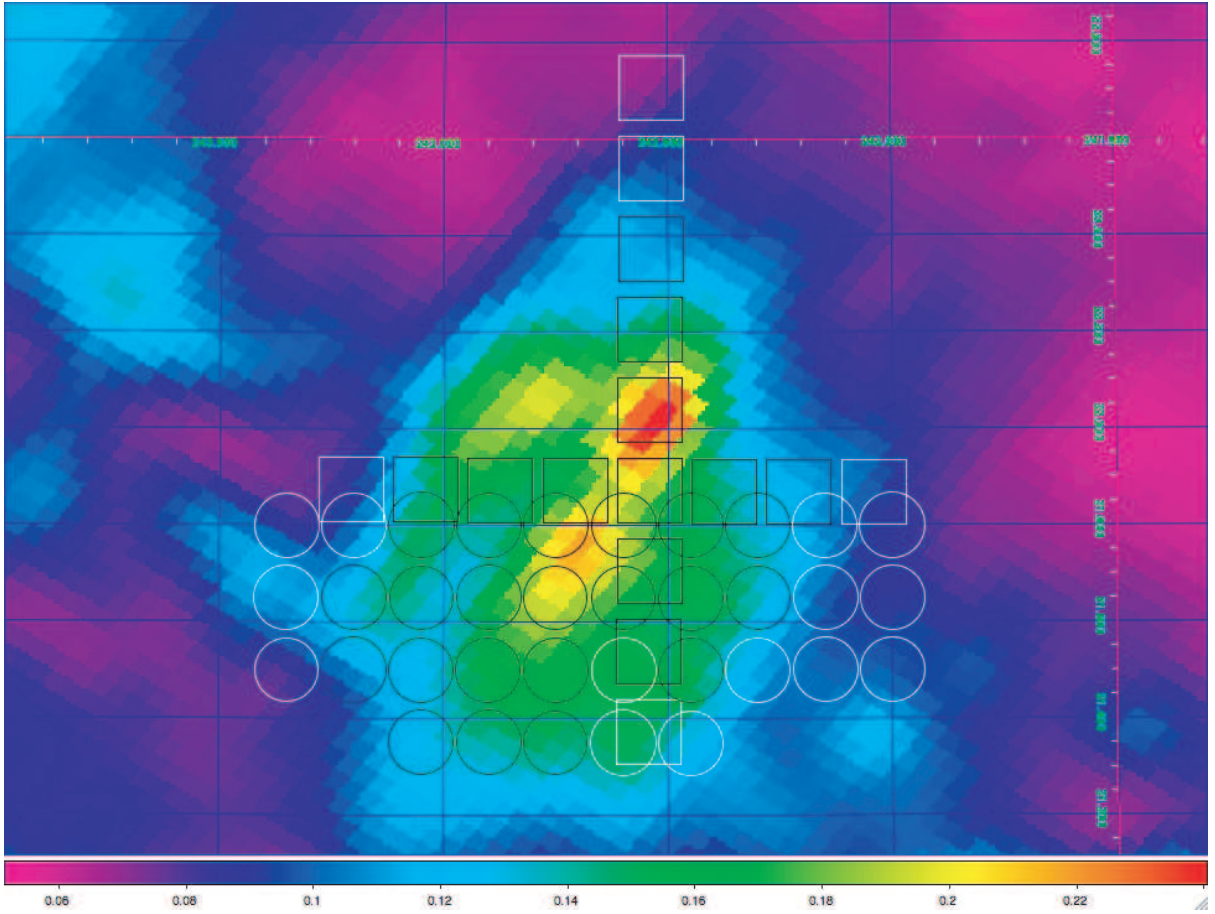}
\caption{E(B-V) map of MBM40 from the Schlegel, Finkbeiner, $\&$ Davis (1998), overlaid with OH observations of the 1667 MHz line from this work (circles) and from \citet[][squares]{wen07}.  Detections are represented by black symbols, non detections by white symbols. See Table~\ref{tab:1} for line parameters. The intensity scale is in MJy/ster.}\label{fig:oh-pos}
\end{center}
\end{figure}

Due to the long integration times needed to detect OH in low extinction regions, data from this work were combined with similar observations done by \citet{wen07} in order to compile a larger data set for analysis. Figure~\ref{fig:oh-pos} shows the SFD dust map for MBM40 overlaid with the positions of observations from this work (circles) and from \citet{wen07} (squares).  The earlier observations were conducted during the months of July and August in 2006 using the GBT in position switching mode, with the same basic setup and ``off'' position criteria used in this work. The combined data resulted in 3 detections out of 18 in the periphery region, 21 out of 25 in the envelope and 8 out of 8 in the core regions. The results of the observations for the 1667 MHz transition are given in Table~\ref{tab:1} where the first two columns give the R.A. and Dec of each observation in J2000 coordinates, while the next three columns list the peak antenna temperature, the LSR centroid velocity, and the FWHM from a Gaussian fit to the profile. Column six tabulates W(OH)$\footnote{W(OH) = $\int$ T$\mathrm{_{R}}$ dv with units of K km s$\mathrm{^{-1}}$.}$; but  if only one number is tabulated then that is the 1-$\sigma$ upper limit of that specific observation. This upper limit is calculated by using the rms value of that specific observation as the temperature in conjunction with the average FWHM from all observations.  The last column lists E(B-V) from the SFD database in magnitudes. The uncertainty in this quantity is $\sim$10$\%$~\citep{SFD98}.

\section{Analysis}

\subsection{W(OH) - E(B-V) Relation}

\begin{figure}
\begin{center}
\includegraphics[width=160mm]{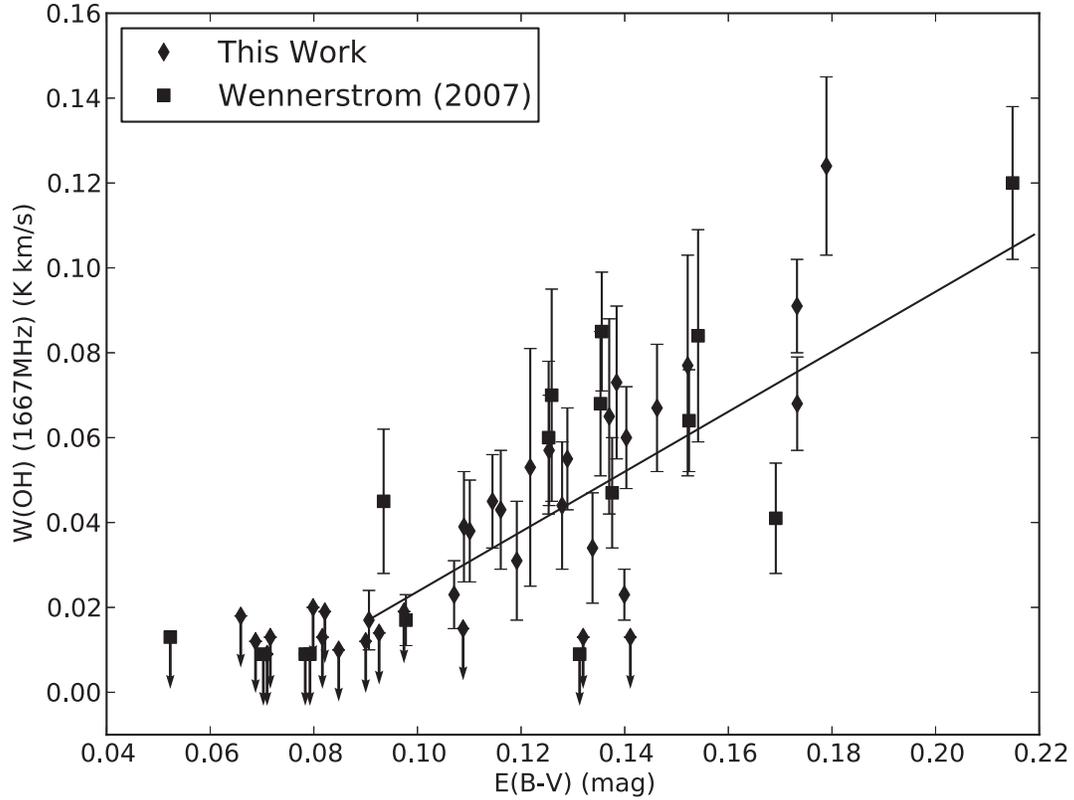}
\caption{W(OH)$_{1667}$ in units of ($\mathrm{K\ km\  s^{-1}}$) versus E(B-V)$\mathrm{_{H_2}}$ in (mag) for MBM40. The arrows represent upper limits for W(OH). The least squares fit applies only to the points that are not upper limits. The resolution of the OH data and the E(B-V) data from \citet{SFD98} is nearly identical (6.2' vs. 6.1'), and the uncertainty in E(B-V) is 10\%.} \label{fig:wohebv}
\end{center}
\end{figure}

In Figure~\ref{fig:wohebv} all the W(OH) detections from Table \ref{tab:1} are plotted versus their associated E(B-V) value from the SFD database, and a line is fit to the data. Non-detections are also plotted as 1-$\sigma$ upper limits. The beam size of the OH observations (6.2') is nearly identical to the resolution of the SFD E(B-V) data (6.1'). The E(B-V) value has been corrected for the dust associated with atomic hydrogen in MBM40. This was done using HI data from \citet{gir94} to estimate N(HI) over a 3 km s$^{-1}$ range centered on the cloud CO velocity.  \citet{shore03} showed that HI emission over this range is directly associated with the molecular cloud. The N(HI) value associated with the cloud is converted to E(B-V) using the~\citet{BSD78} conversion factor. The HI correction only slightly changes the overall E(B-V) value, on average 0.018 mag, because most of the cloud has relatively low N(HI) values $\sim$1x10$^{20}$ (cm$^{-2}$) compared to N(H$_2$) \citep[see][]{RJC05}. We will refer to that portion of the overall E(B-V) that is associated with molecular gas as E(B-V)$\mathrm{_{H_2}}$.  The linear relationship between W(OH)$_{1667}$ and E(B-V)$\mathrm{_{H_2}}$ found using a least squares fit is

\begin{equation}
\label{eq:ebvoh}
\mathrm{W(OH)_{1667}\ =\ (0.71 \pm 0.08) E(B-V)_{H_2}\ -\ 0.05 \pm 0.01\ \ (K\ km\ s^{-1})} 
\end{equation}

\noindent
where E(B-V)$\mathrm{_{H_2}}$ is in magnitudes. The best fit line applies only to positions 
with detections in W(OH) and has a correlation coefficient of 0.78. Positions where W(OH) 
was not detected are labeled with arrows representing upper limits. A survival analysis was 
done including the non-detections to determine if a linear relationship could be assumed 
including the left-censored data. Including  both censored and non-censored data a p value of 
1.1 x 10$\mathrm{^{-5}}$ was obtained indicating that the variables are dependent on each other.
Treating the OH upper limits as part of a bivariate censored problem, the Buckley-James method
and the ASURV\footnote{We used ASURV Rev. 1.2 (\citep{lavalley}) which implements
the methods presented by ]\citep{isobe}.} software package from Penn State University
were used to calculate the linear relationship between all the W(OH)$_{1667}$ and E(B-V) data: 

\begin{equation}
\label{eq:ebvoh_censor}
\mathrm{W(OH)_{1667}\ =\ (0.54 \pm 0.12) E(B-V)_{H_2}\ -\ 0.01 \ \ (K\ km\ s^{-1})}
\end{equation}

Since there are no heating sources in MBM40, the SFD dust map of  the region is very similar to the IRIS 100 $\mu$m map. Thus, a linear relationship will also exist between W(OH) and the 100 $\mu$m radiance, as was found by ~\citet{barr10b} and \citet{gros90} for other translucent clouds. If we convert E(B-V)$\mathrm{_{H_2}}$ to visual extinction using the standard value for the optical total to selective extinction, and W(OH) to N(OH), we derive the $\mathrm{OH/H_2}$ abundance (see section~\ref{sec:ohmass}). The slope of the relationship shown in Figure~\ref{fig:wohebv} falls squarely on the median value from a compilation of primarily translucent lines of sight by~\citet{crut79} (N(OH)/A$\mathrm{_V}$ $\approx$ 8 x 10$\mathrm{^{13}\ cm^{-2}\ mag^{-1}}$).

Using the SFD dust map, E(B-V)$\mathrm{_{H_2}}$ values were found for all the SFD data points 
located in the three regions defined in Section 2, and from that, average E(B-V)$\mathrm{_{H_2}}$ 
values were calculated. We then used the relationship in Equation~\ref{eq:ebvoh_censor} to calculate an 
average W(OH) for the three extinction regimes (see Table~\ref{tab:ebvwoh}). From this average W(OH) value, the area covered by each region, and the ratio of detections to observations, the total mass of OH in the cloud, M$\mathrm{_{OH}}$, can be calculated, as described in the next section.

\subsection{The OH Column Density and Mass}

To obtain a mass for MBM40 we need the distance to the cloud.  \citet{wel89} used echelle spectra near the Na I D lines to set an upper limit on the distance of  $\le$ 140 pc. In 1993, Penprase revised the distance of MBM40 to between 60 and 290 pc using Na I absorption of stars behind the cloud, and to 90 $<$ d $<$ 150 pc by using CH observations. This  improved estimate over the work by \citet{wel89} was due to additional foreground stars used to obtain a lower limit to the distance of MBM40. For this work we have adopted a distance of 120 pc which allows us to calculate the cloud OH mass using: 
 \begin{equation}
\label{eq:mass-oh}
\mathrm{M_{OH} = 4.14 \times 10^{-23}\ \ N(OH)\ \ \Omega\ \ d^{2}\ \ \ (M_{\sun})} 
\end{equation}
where N(OH) is the average column density in cm$^{-2}$ over the region,  $\Omega$ is the solid 
angle covered by the region in question in square degrees, and d is the distance to the cloud in pc. For the column density, we use the formulation described by \citet{W81}:
\begin{equation}
\label{eq:noh}
\mathrm{N(OH)\ =\ \ 2.39\ x\ 10^{14}\ T_{ex}\ \ \Delta v\ \ \tau\ \ \ (cm^{-2})}
\end{equation}
where T$\mathrm{_{ex}}$ is the excitation temperature of the transition, $\Delta$v is the FWHM of the line, and $\tau$ is the optical depth.  
In the optically thin approximation, the radiation temperature is related to optical depth by:
\begin{equation}
\label{eq:ta}
\mathrm{T_{R}\ =\ (T_{ex}\ -\ T_{BG})\ \tau_{\nu}\ \ (K)}
\end{equation} 
where $\mathrm{T_{BG}}$ is the background temperature at 18 cm: 3.3 K \citep[e.g.][]{barr10b}. Since we do not have an independent estimate of  T$\mathrm{_{ex}}$, we determined N(OH) for three plausible values of T$\mathrm{_{ex}}$: 5, 10, and 20 K~\citep{W81, ll96, har00}. N(OH) values ranging from 9.4 x 10$^{12}$ to 5.8 x 10$^{13}$ cm$^{-2}$ were obtained for the three temperature estimates. Average values of 1.2 x 10$^{13}$, 2.1 x 10$^{13}$, and 2.9 x 10$^{13}$ cm$^{-2}$ for the periphery, 
envelope, and core regions, respectively, were found for T$\mathrm{_{ex}}$ = 10 K, and the mass of OH at that excitation temperature in the periphery, envelope, and core was 1.4 x 10$^{-6}$, 6.5 x 10$^{-6}$, and 3.5 x 10$^{-6}$ M$_{\sun}$, respectively. For the 
other values of T$\mathrm{_{ex}}$, the masses in the regions are shown in Table~\ref{tab:moh}. In all cases, the OH mass in the envelope is greater than that in the core.

\subsection{The OH Abundance and Molecular Mass in the Three Regions}\label{sec:ohmass}

To go from the OH mass to the overall molecular mass of the cloud, we must determine the value of  the OH abundance with respect to H$_2$.  One of the earliest comprehensive astrochemical models of translucent and diffuse molecular clouds with A$\mathrm{_V}$ ranging from 0.1 - 1 mag and T$\mathrm{_K}$ from 50 to 100 K was by \citet{viala86}, who found that the OH abundance with respect to H$_{2}$ varied from  3.8 x 10$^{-9}$ to 9.4 x 10$^{-9}$. Another model by \citet{nbv88} used the translucent cloud along the line of sight towards HD 29647 to calculate the OH abundance with respect to H$_2$, and found that it varied from  2.6 x 10$^{-8}$ to 1.2 x 10$^{-6}$ depending on different heavy element depletions, cosmic ray ionization rates, and depth into the cloud. \citet{vDB86} also calculated comprehensive models of diffuse and translucent clouds with 19 models where $\mathrm{n_{H}}$ ranged from 250 to 1000 cm$^{-3}$ and T$\mathrm{_K}$ ranged from 20 to 100 K, yielding OH abundances ranging from  2.72 x 10$^{-9}$ to 1.7 x 10$^{-7}$. Other OH abundance estimates were obtained for translucent clouds by \citet{vanD90} with values in the range  1.1 x 10$^{-7}$ to 2.3 x 10$^{-7}$ .~\citet{and93} modeled the halo around dark molecular clouds, and obtained values for the OH abundance of $\sim$ 10$^{-7}$. Empirically, \citet{mag88} derived somewhat high OH abundances in the range 2 x 10$^{-7}$ to 4 x 10$^{-6}$ for a small sample of translucent clouds, and \citet{wes10} obtained a value of 1.05 $\pm$ 0.24 x 10$^{-7}$ from OH absorption line measurements of 5 translucent lines of sight; virtually identical to the analysis by \cite{ll02}. Because past studies have indicated that the OH abundance in lower density molecular regions can vary widely (4 x 10$^{-9}$ to 1 x 10$^{-6}$), to determine the OH abundance in MBM40 we employ E(B-V) as a surrogate for N(H$_2$), and use our observations to provide N(OH). In this manner a direct estimate of the OH/H$_2$ abundance in MBM40 can be obtained.

\subsubsection{The E(B-V) Method}

We obtain average OH/$\mathrm{H_{2}}$ abundances for the 3 regions by using the well-established relationship between H$\mathrm{_{Total}}$ = 2N(H$_2$) + N(HI) and E(B-V) to determine $\mathrm{H_{2}}$~\citep[e.g.][]{BSD78}. The E(B-V) values used were found by averaging the SFD E(B-V) values for each region. 
We can estimate N(HI) for the lines of sight that we observed by using the HI map of MBM40 from~\citet{gir94}.  In this manner we can determine N(H$_2$) and, 
with the results of the previous section, the 
OH/$\mathrm{H_{2}}$ abundance (see Table 4). The periphery, envelope, and the core have average values 
for the OH/H$_2$ abundance 3.0 x $\mathrm{10^{-7}}$, 1.1 x $\mathrm{10^{-6}}$, and 1.2 x $\mathrm{10^{-6}}$, respectively. We will use these values to go from the OH mass to the molecular mass.

\subsubsection{The Total Molecular Mass of the Core, Envelope, and Periphery}
Using the total OH mass of the cloud derived in Table~\ref{tab:moh} and the OH/$\mathrm{H_{2}}$  
abundance ratios from the previous section, we compute the total 
molecular mass of MBM40 for each region (see Table~\ref{tab:mh2}).  
Our results indicate that most of the mass of MBM 40 is contained in the envelope and periphery
regions.
The core may contain as little as 20\% of the molecular mass of the cloud. 
The mass estimate for the core is actually an upper limit, since observations 
of the core in an optically thin line perforce include foreground and background 
envelope and periphery regions. For at least this cloud there is significant 
molecular gas contained outside of the regions mapped in CO using traditional 
``on the fly'' mapping methods. If these results can be extrapolated to other 
clouds, then traditional CO(1-0) mapping of molecular clouds (i.e., with typical rms 
levels of 0.1 K) may not account for the bulk of the overall cloud molecular mass. 
This would be an important counterpoint to recent claims by \citet{grenier05} 
that half the molecular mass of the Galaxy may be ``dark'' in the sense of 
not traceable by radio spectroscopic means when CO and HI 21cm observations are 
considered.  Our estimates of the M$\mathrm{_{H_2}}$ for the core region of MBM40 compares  favorably with that of \citet{RJC05} who derived a mass of 11 M$_{\sun}$ from CO(1-0) observations.

\section{Conclusions}

High-sensitivity observations of the 1667 MHz OH line in the translucent cloud MBM40 enabled estimates of the molecular gas mass in its core, envelope, and periphery.  Our focus was the low extinction regions (E(B-V) $\le$ 0.17 mag) surrounding the CO-defined core region. From our analysis, we find that nearly half of the cloud's molecular mass may reside in the envelope region surrounding the core, and we find that nearly a third of the molecular gas may reside in the periphery regions of the cloud where E(B-V) is lower than 0.12 mag (equivalent to A$\mathrm{_V}$ $<$ 0.4 mag).  The core emission contains contributions from foreground and  background gas in the envelope and periphery of the cloud.  We have not attempted to correct the core mass for this so our results likely overestimate the contribution of the core to the overall cloud mass.
The total molecular mass derived from our OH data is 17 M$_{\sun}$ which compares favorably with the mass derived by \citet{RJC05} from CO observations. A linear relationship between W(OH) and E(B-V) exists for this cloud underscoring the efficacy of OH 1667 MHz observations in tracing low density molecular gas. OH emission is detected down to E(B-V) level of 0.1 mag, corresponding to $\mathrm{A_V}$ $\sim$ 0.3 mag. The OH 1667 MHz line is similar to the CO(1-0) line and the CH 3335 MHz line in tracing molecular emission at very low extinctions.

Our results confirm that there is significant mass in the regions surrounding 
the CO core of this translucent cloud. These regions are often at extinctions 
that were previously thought to be too low to contain enough molecular gas column 
density to be detected by radio spectroscopic emission lines. Results similar to 
ours have been established using CO data for the Taurus clouds by~\citet{gold08} 
and by Liszt and Pety (2012) for diffuse clouds. If OH does trace the envelope 
regions of translucent clouds, in general, then extensive high-sensitivity OH 
surveys of nearby molecular clouds should be undertaken to determine how much mass 
resides outside the traditionally CO-mapped cores. 

This research was partially supported by the NRAO with the Student Observing Support Award (GSSP09-0018). The National Radio Astronomy Observatory is a facility of the National Science Foundation operated under cooperative agreement by Associated Universities, Inc..  This research has made use of the NASA/IPAC Infrared Science Archive, which is operated by the Jet Propulsion Laboratory, California Institute of Technology, under contract with the National Aeronautics and Space Administration, and the ASURV statistical data analysis 
package from Penn State University. We would like to thank Ron Maddalena for help with the GBT observations. We would also like to thank Steve Shore and Harvey Liszt for a critical reading of the manuscript and an anonymous referee for several helpful
suggestions especially regarding survival analysis of censored data.

\bibliographystyle{aa} 
\bibliography{OH-manuscript-AJ}

\begin{thebibliography}{46}
\expandafter\ifx\csname natexlab\endcsname\relax\def\natexlab#1{#1}\fi

\bibitem[{Andersson \& Wannier(1993)}]{and93}
Andersson, B.-G. \& Wannier, P.~G. 1993, ApJ, 402, 585

\bibitem[{Barriault {et~al.}(2010{\natexlab{a}})Barriault, Joncas, Falgarone,
  \& et~al.}]{barr10a}
Barriault, L., Joncas, G., Falgarone, E., \& et~al. 2010{\natexlab{a}}, MNRAS,
  406, 2713

\bibitem[{Barriault {et~al.}(2010{\natexlab{b}})Barriault, Joncas, Lockman, \&
  Martin}]{barr10b}
Barriault, L., Joncas, G., Lockman, F.~J., \& Martin, P.~G. 2010{\natexlab{b}},
  MNRAS, 407, 2645

\bibitem[{Bergin \& Tafalla(2007)}]{bergin07}
Bergin, E.~A. \& Tafalla, M. 2007, ARA$\&$A, 45, 339

\bibitem[{Black \& Dalgarno(1977)}]{bd77}
Black, J.~H. \& Dalgarno, A. 1977, ApJS, 34, 405

\bibitem[{Blitz \& Williams(1999)}]{blitz99}
Blitz, L. \& Williams, J.~P. 1999, {The Orgin of Stars and Planetary Systems.}
  ({Kluwer Academic Publishers})

\bibitem[{Bohlin {et~al.}(1978)Bohlin, Savage, \& Drake}]{BSD78}
Bohlin, R., Savage, B., \& Drake, J. 1978, The Astrophysical Journal, 224, 132

\bibitem[{Chastain(2008)}]{RJC05}
Chastain, R.~J. 2008, PhD thesis, The University of Georgia

\bibitem[{Chastain {et~al.}(2010)Chastain, Cotten, \& Magnani}]{ccm10}
Chastain, R.~J., Cotten, D.~L., \& Magnani, L. 2010, ApJ, 139, 267

\bibitem[{Chastain {et~al.}(2006)Chastain, Shelton, Raley, \&
  Magnani}]{chastain06}
Chastain, R.~J., Shelton, R.~L., Raley, E.~A., \& Magnani, L. 2006, AJ, 132,
  1964

\bibitem[{Cotten \& Magnani(2012)}]{cotten12}
Cotten, D.~L. \& Magnani, L. 2012, in preparation

\bibitem[{Cox {et~al.}(1989)Cox, Walmsley, \& Guesten}]{cox89}
Cox, P., Walmsley, C.~M., \& Guesten, R. 1989, A$\&$A, 209, 382

\bibitem[{Crutcher(1979)}]{crut79}
Crutcher, R.~M. 1979, ApJ, 234, 881

\bibitem[{deVries(1988)}]{dev88}
deVries, H.~W. 1988, PhD thesis, Columbia University.

\bibitem[{Douglas \& Taylor(2007)}]{dt07}
Douglas, K.~A. \& Taylor, A.~R. 2007, ApJ, 659, 426

\bibitem[{Gir {et~al.}(1994)Gir, Blitz, \& Magnani}]{gir94}
Gir, B., Blitz, L., \& Magnani, L. 1994, ApJ, 434, 162

\bibitem[{Goldsmith {et~al.}(2008)Goldsmith, Heyer, Narayanan, Snell, Li, \&
  Brunt}]{gold08}
Goldsmith, P.~F., Heyer, M., Narayanan, G., {et~al.} 2008, ApJ, 680, 428

\bibitem[{Grenier {et~al.}(2005)Grenier, Casandjian, \& Terrier}]{grenier05}
Grenier, I.~A., Casandjian, J.~M., \& Terrier, R. 2005, Science, 307, 1292

\bibitem[{Grossman {et~al.}(1990)Grossman, Meyerdierks, Mebold, \&
  Heitausen}]{gros90}
Grossman, V., Meyerdierks, H., Mebold, U., \& Heitausen, A. 1990, A$\&$A, 240,
  400

\bibitem[{Harju {et~al.}(2000)Harju, Winnberg, \& A.}]{har00}
Harju, J., Winnberg, A., \& A., W. J.~G. 2000, A$\&$A, 353, 1065

\bibitem[{Hearty {et~al.}(1999)Hearty, Magnani, Caillault, Neuh$\ddot{a}user,
  Schmitt, \& Stauffer}]{hearty99}
Hearty, T., Magnani, L., Caillault, J.-P., {et~al.} 1999, A$\&$A, 341, 163

\bibitem[{Hollenbach \& Tielens(1997)}]{ht97}
Hollenbach, D.~J. \& Tielens, A. G. G.~M. 1997, ARA$\&$A, 35, 179

\bibitem[{Ingalls {et~al.}(1997)Ingalls, Chamberlin, Bania, Jackson, Lane, \&
  Stark}]{ing97}
Ingalls, J.~G., Chamberlin, R.~A., Bania, T.~M., {et~al.} 1997, ApJ, 479, 296

\bibitem[{Isobe {et~al.}(1986)Isobe, Feigelsen, \& Nelson}]{isobe}
Isobe, T., Feigelsen, E., \& Nelson, P. 1986, ApJ, 306, 490

\bibitem[{LaValley {et~al.}(1992)LaValley, Isobe, \& Feigelsen}]{lavalley}
LaValley, M., Isobe, T., \& Feigelsen, E. 1992, BAAS, 24, 839

\bibitem[{Liszt \& Lucas(1996)}]{ll96}
Liszt, H. \& Lucas, R. 1996, A$\&$A, 314, 917

\bibitem[{Liszt \& Lucas(2002)}]{ll02}
Liszt, H. \& Lucas, R. 2002, A$\&$A, 384, 1054

\bibitem[{Maddalena(2009)}]{Madd09}
Maddalena, R. 2009, {The Performance of the GBT A Guide for Planning
  Observations},
  \texttt{http://www.gb.nrao.edu/~rmaddale/GBT/ReceiverPerformance/PlaningObse%
rvations.htm}

\bibitem[{Magnani {et~al.}(1985)Magnani, Blitz, \& Mundy}]{mbm85}
Magnani, L., Blitz, L., \& Mundy, L. 1985, ApJ, 295, 402

\bibitem[{Magnani {et~al.}(1988)Magnani, Blitz, \& Wouterloot}]{mag88}
Magnani, L., Blitz, L., \& Wouterloot, J. 1988, ApJ, 326, 909

\bibitem[{Magnani \& Onello(1993)}]{mo93}
Magnani, L. \& Onello, J.~S. 1993, ApJ, 408, 559

\bibitem[{Magnani \& Siskind(1990)}]{magsis90}
Magnani, L. \& Siskind, L. 1990, ApJ, 359, 355

\bibitem[{McGehee(2008)}]{mcgehee}
McGehee, P.~M. 2008, {Handbook of Star Forming Regions, Volume II: The Southern
  Sky ASP Monograph Publications, Vol. 5} (B. Reipurth)

\bibitem[{Miville-Desch$\mathrm{\hat{e}}$nes \& Lagache(2005)}]{miv05}
Miville-Desch$\mathrm{\hat{e}}$nes, M.-A. \& Lagache, G. 2005, ApJS, 157, 302

\bibitem[{Nercessian {et~al.}(1988)Nercessian, Benayoun, \& Viala}]{nbv88}
Nercessian, E., Benayoun, J.~J., \& Viala, Y.~P. 1988, A$\&$A, 195, 245

\bibitem[{Schlegel {et~al.}(1998)Schlegel, Finkbeiner, \& Davis}]{SFD98}
Schlegel, D.~J., Finkbeiner, D.~P., \& Davis, M. 1998, ApJ, 500, 525

\bibitem[{Shore {et~al.}(2003)Shore, Magnani, LaRosa, \& McCarthy}]{shore03}
Shore, S.~N., Magnani, L., LaRosa, T.~N., \& McCarthy, M.~N. 2003, ApJ, 593,
  413

\bibitem[{van Dishoeck(1990)}]{vanD90}
van Dishoeck, E. 1990, PASP, 12, 207

\bibitem[{van Dishoeck \& Black(1986)}]{vDB86}
van Dishoeck, E.~F. \& Black, J.~H. 1986, ApJS, 62, 109

\bibitem[{van Dishoeck \& Black(1988)}]{vDB88}
van Dishoeck, E.~F. \& Black, J.~H. 1988, ApJ, 334, 771

\bibitem[{Viala(1986)}]{viala86}
Viala, Y.~P. 1986, A$\&$AS, 64, 391

\bibitem[{Wannier {et~al.}(1993)Wannier, Andersson, Federman, Lewis, Viala, \&
  Shaya}]{wan93}
Wannier, P.~G., Andersson, B.-G., Federman, S.~R., {et~al.} 1993, ApJ, 407, 163

\bibitem[{Welty {et~al.}(1989)Welty, Hobbs, Blitz, \& Penprase}]{wel89}
Welty, D.~E., Hobbs, L.~M., Blitz, L., \& Penprase, B.~E. 1989, ApJ, 346, 232

\bibitem[{Wennerstrom(2007)}]{wen07}
Wennerstrom, E. 2007, Master's thesis, The University of Georgia

\bibitem[{Weselak {et~al.}(2010)Weselak, Galazutidinov, Beletsky, \&
  Krelowski}]{wes10}
Weselak, T., Galazutidinov, G.~A., Beletsky, Y., \& Krelowski, J. 2010, MNRAS,
  402, 1991

\bibitem[{Wouterloot(1981)}]{W81}
Wouterloot, J. 1981, PhD thesis, The University of Leiden: Leiden.

\end{thebibliography}

\begin{deluxetable}{l  c    c   c   c   c   c}
%\tabletypesize{\scriptsize}
\tablecaption{OH 1667 MHz observations  in MBM40.\label{tab:1}}
\tablewidth{0pt}
\tablehead{
\colhead{RA (2000)} & \colhead{Dec (2000)} & \colhead{$\mathrm{T_{A}}$} &
\colhead{$\mathrm{v_{LSR}}$} & \colhead{$\mathrm{\Delta v}$} & \colhead{W(OH)\tablenotemark{a}} &  \colhead{E(B-V)$\mathrm{_{H_2}}$\tablenotemark{b}}\\
\colhead{(deg)} & \colhead{(deg)} & \colhead{(K)} &
\colhead{(km s$^{-1}$)} & \colhead{(km s$^{-1}$)} & \colhead{(K km s$^{-1}$)} &  \colhead{(mag)}
}
\startdata
243.35 &       21.80 &                    &        &                            &    0.009 &0.071\\
243.20 &       21.80 &                    &        &                            &     0.009&0.085 \\
243.05 &       21.80 & 0.075  +/- 0.016   &  2.94 &       0.64 +/- 0.13 &   0.055 +/- 0.012&0.129 \\
242.90 &       21.80 & 0.070  +/- 0.016   &  3.25 &       0.71 +/- 0.16 &   0.057 +/- 0.013&0.125 \\
242.75 &       21.80 & 0.155  +/- 0.018   &  3.31 &       0.51 +/- 0.06 &   0.091 +/- 0.011&0.173 \\
242.60 &       21.80 & 0.103  +/- 0.017   &  3.34 &       0.58 +/- 0.09 &   0.068 +/- 0.011&0.173 \\
242.45 &       21.80 & 0.052  +/- 0.012   &  3.73 &       0.39 +/- 0.09 &   0.023 +/- 0.006&0.140 \\
242.30 &       21.80 & 0.057  +/- 0.014   &  3.66 &       0.69 +/- 0.17 &   0.045 +/- 0.011&0.115 \\
242.15 &       21.80 &                    &       &                     &             0.012&0.090 \\
242.00 &       21.80 &                    &       &                     &             0.012&0.069 \\
243.35 &       21.65 &                    &       &                     &             0.014&0.093   \\
243.20 &       21.65 & 0.041  +/- 0.013   &  2.59 &       0.81 +/- 0.26 &   0.038 +/- 0.012&0.110 \\
243.05 &       21.65 & 0.032  +/- 0.014   &  3.08 &       0.84 +/- 0.37 &   0.031 +/- 0.014&0.119 \\
242.90 &       21.65 & 0.081  +/- 0.020   &  3.31 &       0.79 +/- 0.19 &   0.073 +/- 0.018&0.138 \\
242.75 &       21.65 & 0.129  +/- 0.022   &  3.31 &       0.84 +/- 0.14 &   0.124 +/- 0.021&0.179 \\
242.60 &       21.65 & 0.043  +/- 0.014   &  3.48 &       1.57 +/- 0.52 &   0.077 +/- 0.026&0.152 \\
242.45 &       21.65 & 0.041  +/- 0.015   &  3.24 &       1.40 +/- 0.50 &   0.065 +/- 0.023&0.137 \\
242.30 &       21.65 & 0.042  +/- 0.014   &  3.29 &       0.90 +/- 0.29 &   0.043 +/- 0.014&0.116 \\
242.15 &       21.65 &                    &       &                     &             0.013&0.082 \\
242.00 &       21.65 &                    &       &                     &             0.013&0.072 \\
243.35 &       21.50 &                    &       &                     &             0.018&0.066 \\
243.20 &       21.50 & 0.033  +/- 0.014   &  2.59 &       0.46 +/- 0.19 &   0.017 +/- 0.007&0.091 \\
243.05 &       21.50 & 0.049  +/- 0.016   &  3.32 &       0.69 +/- 0.23 &   0.039 +/- 0.013&0.109 \\
242.90 &       21.50 & 0.083  +/- 0.016   &  3.32 &       0.63 +/- 0.12 &   0.060 +/- 0.012&0.140 \\
242.75 &       21.50 & 0.075  +/- 0.016   &  3.55 &       0.77 +/- 0.17 &   0.067 +/- 0.015&0.146 \\
242.60 &       21.50 &                    &       &                     &             0.013&0.141 \\
242.45 &       21.50 & 0.037  +/- 0.014   &  3.20 &       0.80 +/- 0.30 &   0.034 +/- 0.013&0.134 \\
242.30 &       21.50 &                    &       &                     &             0.019&0.097 \\
242.15 &       21.50 &                    &       &                     &             0.019&0.082 \\
242.00 &       21.50 &                    &       &                     &             0.020&0.080 \\
243.05 &       21.35 & 0.061  +/- 0.020   &  3.02 &       0.34 +/- 0.11 &   0.023 +/- 0.008&0.107 \\
242.90 &       21.35 & 0.036  +/- 0.019   &  3.42 &       1.30 +/- 0.67 &   0.053 +/- 0.028&0.122 \\
242.75 &       21.35 & 0.051  +/- 0.017   &  3.19 &       0.75 +/- 0.25 &   0.044 +/- 0.015&0.128 \\
242.60 &       21.35 &                    &       &             &                     0.013&0.132 \\
242.45 &       21.35 &                    &        &             &                    0.015&0.109  \\
\hline
\multicolumn{2}{c}{{Wennerstrom (2007) data}}\\
\hline
242.54 &        21.87 & 0.050  +/- 0.016 &  3.25   &  0.71 +/- 0.22&0.041 +/- 0.013& 0.169 \\
242.54 &        22.03 & 0.123  +/- 0.018 &  2.96   &  0.85 +/- 0.12&0.120 +/- 0.018& 0.215  \\
242.53 &        22.20 & 0.107  +/- 0.018 &  3.06   &  0.69 +/- 0.11&0.085 +/- 0.014& 0.136  \\
242.53 &        22.37 & 0.044  +/- 0.017 &  3.47   &  0.88 +/- 0.34&0.045 +/- 0.017& 0.093  \\
242.53 &        22.53 &                  & & &                                0.009& 0.079  \\
242.53 &        22.70 &                  & & &                                0.013& 0.052  \\
242.54 &        21.70 & 0.055  +/- 0.017 &  3.49   &  1.34 +/- 0.40&0.084 +/- 0.025& 0.151  \\
242.54 &        21.53 & 0.040  +/- 0.011 &  3.11   &  1.03 +/- 0.29&0.047 +/- 0.013& 0.138  \\
242.54 &        21.37 &                  & & &                                0.009& 0.131  \\
242.37 &        21.86 & 0.056  +/- 0.017 &  3.65   &  0.93 +/- 0.27&0.060 +/- 0.018& 0.125  \\
242.20 &        21.86 & 0.030  +/- 0.010 &  3.62   &  0.49 +/- 0.16&0.017 +/- 0.006& 0.098  \\
242.04 &        21.86 &                  & & &                                0.009& 0.078  \\
242.70 &        21.87 & 0.087  +/- 0.016 &  3.30   &  0.64 +/- 0.11&0.064 +/- 0.012& 0.152  \\
242.87 &        21.87 & 0.072  +/- 0.018 &  3.03   &  0.82 +/- 0.20&0.068 +/- 0.017& 0.135  \\
243.04 &        21.87 & 0.048  +/- 0.017 &  3.12   &  1.29 +/- 0.46&0.070 +/- 0.025& 0.126  \\
243.20 &        21.87 &        &   &      &                                   0.009& 0.070  \\
\enddata
\tablenotetext{a}{If there is no error associated with the W(OH) entry, then it is a 1-$\sigma$ upper limit.}
\tablenotetext{b}{E(B-V)$\mathrm{_{H_2}}$ is the color excess associated with the molecular gas in the cloud. The contribution from dust associated with HI has been subtracted.}
\end{deluxetable}

\begin{deluxetable}{l  c    c }
%\tabletypesize{\scriptsize}
\tablecaption{Average E(B-V)$\mathrm{_{H_2}}$ and W(OH)$_{1667}$ values for the MBM40 regions (see Section 2). \label{tab:ebvwoh}}
\tablewidth{0pt}
\tablehead{
\colhead{Region} & \colhead{E(B-V)$\mathrm{_{H_2}}$} &  \colhead{W(OH)$_{1667}$} \\
\colhead{} &  \colhead{(mag)} &  \colhead{(K km s$^{-1}$)} 
}
\startdata
Periphery & 0.08 & 0.033 \\ 
Envelope & 0.13 & 0.060\\ 
Core & 0.17 & 0.082\\
\enddata 
\end{deluxetable}

\begin{deluxetable}{l  c    c  c   c}
%\tabletypesize{\scriptsize}
\tablecaption{Total M(OH) for 3 regions in MBM40.  \label{tab:moh}}
\tablewidth{0pt}
\tablehead{
\colhead{Region} & \colhead{detection} & \colhead{$\mathrm{T_{ex}}$ = 5 K} &  \colhead{$\mathrm{T_{ex}}$ = 10 K} &  \colhead{$\mathrm{T_{ex}}$ = 20 K}\\
\colhead{} & \colhead{fraction} & \colhead{($\mathrm{10^{-6} M_{\odot}}$)} &  \colhead{($\mathrm{10^{-6} M_{\odot}}$)} & \colhead{($\mathrm{10^{-6} M_{\odot}}$)} 
}
\startdata
Periphery & 0.17  &  2.8 $\pm$ 3.3  &  1.4 $\pm$ 1.7  &  0.6 $\pm$ 0.7  \\ 
Envelope  & 0.84  & 12.9 $\pm$ 4.2  &  6.5 $\pm$ 2.3  &  5.2 $\pm$ 1.7  \\ 
Core      & 1.0   &  6.9 $\pm$ 1.6  &  3.5 $\pm$ 0.9  &  2.8 $\pm$ 0.7 \\
\enddata
\end{deluxetable}

\begin{deluxetable}{l  c    c  c   c}
%\tabletypesize{\scriptsize}
\tablecaption{OH/$\mathrm{H_2}$ abundance ratio ($\mathrm{10^{-7}}$) for 3 regions in MBM40.  \label{tab:ohh2}}
\tablewidth{0pt}
\tablehead{
\colhead{Region} &  \colhead{$\mathrm{T_{ex}}$ = 5 K} &  \colhead{$\mathrm{T_{ex}}$ = 10 K} &  \colhead{$\mathrm{T_{ex}}$ = 20 K} & \colhead{Average}
}
\startdata
Periphery &  5.6 $\pm$ 3.4  &  2.3 $\pm$ 1.4  &  1.2 $\pm$ 0.7  &  3.0 \\ 
Envelope  & 16.9 $\pm$ 4.7  &  8.5 $\pm$ 2.5  &  6.7 $\pm$ 1.9  & 10.9 \\ 
Core  	  & 18.5 $\pm$ 3.9  &  9.6 $\pm$ 2.0  &  7.2 $\pm$ 1.5  & 11.8 \\
\enddata
\end{deluxetable}

\begin{deluxetable}{l  c    c  c   }
%\tabletypesize{\scriptsize}
\tablecaption{Total M($\mathrm{H_{2})}$ for MBM40.  \label{tab:mh2}}
\tablewidth{0pt}
\tablehead{
\colhead{Region} &  \colhead{Mass}\\
\colhead{} &  \colhead{(M$_{\sun}$)} 
}
\startdata
Periphery & 5.2 $\pm$ 6.3  \\ 
Envelope &  7.6 $\pm$ 2.5 \\ 
Core & 3.8 $\pm$ 0.9   \\
\hline
Total &  16.6 $\pm$ 9.7  \\
\enddata
\end{deluxetable}

\end{document}